High Performance WSe$_2$ Field-Effect Transistors *via* Controlled Formation of In-Plane Heterojunctions

Bilu Liu,[1,†] Yuqiang Ma,[1,†] Anyi Zhang,[1] Liang Chen,[1] Ahmad N. Abbas,[1] Yihang Liu,[1] Chenfei Shen,[1] Haochuan Wan,[1] Chongwu Zhou[1*]

[1] Ming Hsieh Department of Electrical Engineering, University of Southern California, Los Angeles, California 90089, USA

[†] Equal contribution.

[*] Corresponding author. E-mail: chongwuz@usc.edu

*Abstract*

**Monolayer WSe$_2$ is a two dimensional (2D) semiconductor with a direct bandgap, and it has been recently explored as a promising material for electronics and optoelectronics. Low field effect mobility is the main constraint preventing WSe$_2$ from becoming one of the competing channel materials for field-effect transistors (FETs). Recent results have demonstrated that chemical treatments can modify the electrical properties of transition metal dichalcogenides (TMDCs) including MoS$_2$ and WSe$_2$. Here, we report that controlled heating in air significantly improves device performance of WSe$_2$ FETs in terms of on-state currents and field-effect mobilities. Specifically, after heating at optimized conditions, chemical vapor deposition grown monolayer WSe$_2$ FETs showed an average**

**FET mobility of 31 $cm^2 \cdot V^{-1} s^{-1}$ and on/off current ratios up to $5×10^8$. For few-layer $WSe_2$ FETs, after the same treatment applied, we achieved a high mobility up to 92 $cm^2 \cdot V^{-1} s^{-1}$. These values are significantly higher than FETs fabricated using as-grown $WSe_2$ flakes without heating treatment, demonstrating the effectiveness of air heating on the performance improvements of $WSe_2$ FETs. The underlying chemical processes involved during air heating and the formation of in-plane heterojunctions of $WSe_2$ and $WO_{3-x}$, which is believed to be the reason for the improved FET performance, were studied by spectroscopy and transmission electron microscopy. We further demonstrated that by combining air heating method developed in this work with supporting 2D materials on BN substrate, we achieved a noteworthy field effect mobility of 83 $cm^2 \cdot V^{-1} s^{-1}$ for monolayer $WSe_2$ FETs. This work is a step towards controlled modification of the properties of $WSe_2$ and potentially other TMDCs, and may greatly improve device performance for future applications of 2D materials in electronics and optoelectronics.**

Two dimensional (2D) semiconducting transition metal dichalcogenides (TMDCs) have ignited substantial scientific interest in the past few years owning to their unique structures and properties, which satisfy requests of many electronic devices. One of the primary advantages of field effect transistors (FETs) based on atomically thin TMDCs is the absence of dangling bonds on the top and bottom surfaces of these materials, which prevent devices from performance degradation due to interface states.[1-4] In addition, the ultrathin body of TMDCs can result in a strong electrostatic gating of these materials, making them attractive for electronics and optoelectronics.[5] Another advantage for TMDCs is that by tuning compositions, number of layers, and strains, scientists can modify the electronic properties of these materials.[6-15] Accordingly, direct or indirect bandgap materials, and n-type, p-type or ambipolar FET devices, have been fabricated.[2, 8, 15-20] The enhanced spin-orbit coupling in TMDCs also makes it possible to use spin and valley information of carriers for novel concept devices.[21] Meanwhile, investigations of monolayer TMDCs have also indicated new physics like many-body quasiparticles including trions and biexcitons, which have no analogue in bulk semiconductors.[22] Thanks to these extraordinary merits, devices fabricated from either mechanically-exfoliated or vapor-phase-grown TMDCs have been proven to show interesting electronic, optical, catalytic, and energy storage properties.[2, 13, 17, 23, 24] Among all TMDCs, semiconducting $MoS_2$ is the one which has received the most attention while research interest in $WSe_2$ has shown rapid increase recently.[4, 13] Compared with monolayer $MoS_2$, monolayer $WSe_2$ possesses a smaller bandgap (~1.6 eV in monolayer $WSe_2$ and ~1.8 eV in monolayer $MoS_2$), and in average, the field effect mobilities are higher in $WSe_2$ FETs than in $MoS_2$ FETs. Recent experiments have also showed that both n-type FETs,[18] p-type FETs,[16] as well as complementary inverters[17] were fabricated based on $WSe_2$, which has been proven to be difficult in $MoS_2$ due to Fermi level pinning effects. Therefore, from the aforementioned comparisons, WSe$_2$ appears promising for future FET applications.

Because of large amount of surface atoms, TMDCs and other 2D materials have shown to be very susceptible to environments.[25-28] This feature benefits their applications in chemical and biomedicine sensing, but on the other hand, raises a crucial concern about stability of TMDC devices. Recent results have indicated that TMDCs can be oxidized at moderate temperatures in oxygen or ozone environment, leading to the formation of metal oxides which have distinct electronic properties with TMDC themselves.[29-32] Therefore, the stability of TMDC devices in ambient air is of particular importance concerning the practical use of these materials. Moreover, recent reports also indicate that properties of TMDCs are tunable, by means of chemical treatments such as ion intercalation. These facts make it also interesting to explore whether we can take advantages of such treatments to modify the properties of TMDCs and further to improve their FET performance, because low mobility is one of the key limiting factors that hinder the applications of TDMC FETs currently. In this work, we demonstrated such a simple yet effective air heating method which significantly improves device performance of WSe$_2$ FETs *via* controlled formation of in-plane heterojunctions of WSe$_2$ and WO$_{3-x}$.

**RESULTS AND DISCUSSION**

We first show the effectiveness of air heating in improving FET performance of chemical vapor deposition (CVD) grown monolayer WSe$_2$. The CVD-grown WSe$_2$ flakes used in these experiments were synthesized at 950 °C and were monolayers, as revealed by atomic force microscopy (AFM), Raman, and photoluminescence (PL) spectroscopy characterization (Supplementary Information Figure S1). As mentioned above, monolayer materials are very susceptible to their environments, which may have non-trivial influences on their electronic properties. Surprisingly, we found that the on-state current of back-gated monolayer WSe$_2$ FETs increased by 8-fold, field effect mobility increased by 5-fold, and current on/off ratios increased by 10-fold, after leaving the device in ambient air at room temperature for an extended duration of 9 months (Supplementary Figure S2). This phenomenon, *i.e.*, improved FET performance after leaving devices in air, was observed for all the devices (>30) we fabricated, showing the results were highly reproducible. Then an important question raised is whether this process could be accelerated by certain means. It is quite rational to think of raising the temperature while leaving the devices in air, *i.e.*, heating the WSe$_2$ FETs in air at elevated temperatures. We have conducted systematical studies of air heating on several WSe$_2$ FETs under different heating temperatures, to find out the optimal temperature. Figure 1a shows the transfer characteristics ($I_{ds}$-$V_g$) of a typical monolayer WSe$_2$ FET in as-fabricated form, as well as after heating in air at 50, 100, 150, 200, and 250 °C. As can be clearly discerned, the on-state currents keep increasing when the FET was sequentially heated and reached a maximum of 12 µA after 250 °C treatment. Noticeably, compared to the as-fabricated FET, the on-current of the FET heated at 250 °C increased by 4,000 times. Figure 1b shows $I_{ds}$-$V_g$ curves in a semi-log scale of three typical devices after 250 °C air heating, all possessing high on/off ratios above $10^7$, which are significantly greater than the as-fabricated FETs. At different heating temperatures, these devices exhibited a substantial increase in the normalized on-state current ($I_{on}$/W, here W is the channel width of FETs), as shown in Figure 1c. All of them exhibited a maximum on-state current after 250 °C air heating. For the monolayer $WSe_2$ FETs heated in air at 300 °C, we found that the on-state currents significantly decreased (Supplementary Information Figure S3). After further heating to 400 °C, all devices behaved as open circuits with noise level current and showed no gate dependence. The field-effect mobility *versus* heating conditions of these devices show very similar trends with that of the on-current, *i.e.*, the FETs possess maximum field-effect mobilities after 250 °C air heating (Figure 1d). Additionally, we observed that the subthreshold voltage ($V_{th}$) of these devices gradually shifted to more positive voltages as the air heating temperature increased (Figure 1e). We note that all the measured FETs show low off-state current of ~1 pA and high current on/off ratios up to $5 \times 10^8$ was obtained. Figure 1f shows the field-effect mobility statistics of as-fabricated monolayer $WSe_2$ FETs and the same set of devices after heating in air at optimized temperature of 250 °C. Noticeably, the maximum room temperature field-effect mobility reaches 51 $cm^2$/V s for the air heated monolayer $WSe_2$ FETs. After 250 °C air treatment, the average effective mobility of these CVD grown monolayer $WSe_2$ is 31 $cm^2$/V s, which compares favorably to recent reports of monolayer TMDC based FETs reported recently.[8, 33-35] We also noted that these air heated devices show relatively small variations of mobility, compared to other recently results.[35, 36] Moreover, we found that after heating in air at elevated temperatures, the WSe$_2$ FETs became very stable and we did not observe any noticeable change in device characteristics when leaving them in ambient air at room temperature for months. This result suggests that besides the improved FET performance, air heating at high temperatures can also suppress the environmental susceptibility of WSe$_2$ devices at normal room temperature.

To examine the generality of the above air heating method in improving device performance of WSe$_2$ FETs, we have further applied this method to transistors made of few-layer WSe$_2$ flakes, and we demonstrated that the same method also worked for few-layer WSe$_2$ FETs. The few-layer WSe$_2$ samples were grown by physical vapor deposition (PVD) method using WSe$_2$ powders as source materials (see Methods section for the details of the PVD growth and Supplementary Information Figure S4 for the characterization of PVD-grown WSe$_2$ samples). Figure 2a shows $I_{ds}$-$V_g$ curves of a typical few-layer WSe$_2$ FET in the as-fabricated state and after air heating at 150 and 200 °C. As can be clearly seen, similar improvement of on-state current after device heating in air was observed. This observation suggests the generality of the above developed air heating method in improving device performance of both monolayer and few-layer WSe$_2$ samples grown by different methods. The field-effect mobility distributions of as-fabricated few-layer WSe$_2$ FETs and the devices after heating in air at 200 °C were plotted in Figure 2b. Again, one can see that the air heated few layer WSe$_2$ FETs possess much higher mobilities than the as-fabricated FETs, with a maximum mobility of 92 cm$^2$/V·s. We noted that for few-layer WSe$_2$, the optimized heating temperature is 200 °C. After further increasing the temperature to 250 °C, even though the on-state current kept increasing, the on/off ratios dropped significantly for few layer WSe$_2$ FETs (Supplementary Information Figure S5). The reason for such different behaviors between monolayer and few-layer WSe$_2$ FETs will be discussed later.

It is important to understand what happened during air heating, and the mechanism behind the improvement of device performance for WSe$_2$ FETs heated in air. To study the chemistry involved during air heating process, we performed both macro- and micro-scale characterization to investigate the structures and properties of as-prepared and air heated WSe$_2$ samples. For these experiments, we used PVD-grown few-layer WSe$_2$ since PVD synthesis does not require the use of WO$_3$ as a precursor during WSe$_2$ growth (see Methods), which can exclude the influence of oxide in source materials in subsequent spectroscopy analysis. X-ray photoemission spectroscopy (XPS) measurements on as-prepared WSe$_2$ show two major peaks at ~34.7 eV and 32.5 eV, corresponding to the W$^{4+}$ 4f$_{5/2}$ and W$^{4+}$ 4f$_{7/2}$ in WSe$_2$, respectively.[32, 37] A small peak at 30.1 eV was also observed, corresponding to pure W which may be a decomposition product of WSe$_2$ precursor. Small peaks at ~38.2 eV and 36.1 eV, which can be assigned to W$^{6+}$ 4f$_{5/2}$ and W$^{6+}$ 4f$_{7/2}$ in WO$_3$,[37] were also observed. The existence of weak oxide peaks in as-grown PVD WSe$_2$ may due to air exposure of the samples before loading them into XPS chamber. Noticeably, one can see that as the air heating temperature increases, the intensity ratios of W$^{6+}$/W$^{4+}$ keep increasing. These results indicate gradually incorporation of oxygen into WSe$_2$ and the formation of WO$_3$ after air heating. We also conducted X-ray diffraction (XRD, Figure 3b) measurements on as-grown and air heated WSe$_2$ samples. The appearance of new diffraction peaks at 32.9, 61.4, 65.6, 66.3° suggests the formation of WO$_3$, in good accordance with the XPS results in Figure 3a. Observations from micro-Raman performed on air heated WSe$_2$ flakes again show the formation of tungsten oxides at 695 and 803 cm$^{-1}$ (Figure 3c), which are very close to the Raman peaks of WO$_3$. Certain red shifts for both peaks can be caused by the existence of oxygen deficiency, *i.e.*, the formation of non-stiochemistry oxidized species WO$_{3-x}$.[38] Collectively, the above XPS, XRD, and Raman spectroscopic results suggest the formation of WO$_{3-x}$ in air heated WSe$_2$ flakes, which is in consistent with recent studies on the oxidation of MoS$_2$ and WSe$_2$ flakes.[30, 39]

We also performed high-resolution transmission electron microscopy (HRTEM) studies on the air heated WSe$_2$, and found the formation of in-plane heterojunctions of WSe$_2$ and WO$_{3-x}$ in air heated WSe$_2$ flakes (Figure 4a-d). From the fast Fourier transforms (FFTs, Figure 4b-c) of the selected regions in Figure 4a, one can clearly identify both hexagonal WSe$_2$ as well as cubic WO$_{3-x}$ phases. Figure 4d clearly shows the interface between WSe$_2$ and WO$_{3-x}$ domains. The two phases form so-called quilted in-plane heterostructures, as schematically shown in Figure 4e. The sizes of the newly formed cubic $WO_{3-x}$ domains are in the range of a few to a few tens of nanometers. This kind of quilted heterostructures was also found in liquid-exfoliated $MoS_2$ flakes very recently.[40]

Then the question turns to be why the formation of $WSe_2/WO_{3-x}$ in-plane heterojunctions improved device performance of $WSe_2$ FETs. We speculate that there are two effects that may be responsible for the improved device performance of air heated $WSe_2$ FETs. First, it has been well documented that non-stiochemistry tungsten oxide ($WO_{3-x}$) is a highly conductive species. For example, Liu *et al.* have shown that partial oxidation of $WSe_2$ resulted in the formation of $WO_{3-x}$, which is highly conductive with a sheet resistance of ~$10^5$ ohm/sq, while $WSe_2$ possess a three orders of magnitude higher sheet resistance of ~$10^8$ ohm/sq.[30] Therefore, the effective channel lengths of FETs would be reduced *via* the formation of highly conductive $WO_{3-x}$ domains in $WSe_2$. This will in turn result in increase of device on-state current and field-effect mobility. Second, due to the work function mismatch between $WSe_2$ and $WO_{3-x}$, $WO_{3-x}$ can dope $WSe_2$ in a p-type manner and increase hole concentrations in the device, which could increase the on-state current of p-type FETs as well. Indeed, $WO_{3-x}$ has been frequently used as a hole injection material in organic light emission devices, and p-type doping of $WSe_2$ by $WO_{3-x}$ has been recently reported by Tsukagoshi's group.[32] In addition, such p-type doping of $WSe_2$ by $WO_{3-x}$ would lead to a change of $V_{th}$ of FETs towards more positive voltages, consistent with the experimental results shown in Figure 1e. Since $WSe_2$ and $WO_{3-x}$ have different crystal structures and lattice constants, in-plane strain would be introduced during the $WSe_2$ to $WO_{3-x}$ conversion process. The lattice mismatch may result in local physical gaps between $WSe_2$ and $WO_{3-x}$ domains at some points. This situation happens only at high enough temperatures when sufficient amount of $WO_{3-x}$ domains formed and comparable to $WSe_2$ domains. In this case, on-state currents for FETs based on monolayer $WSe_2$ will drop dramatically (as shown in Supplementary Information Figure S3), and eventually the devices can end up with open circuits. However, for multilayer $WSe_2$ flakes, the situation is a bit different as they are more tolerant to the in-plane cracking, since the devices can still be conductive due to charge transport in vertical direction, in addition to the in-plane transport. Therefore, this would lead to a highly conductive thick $WSe_2$ flakes after air heating at elevated temperatures, consistent with our experiments (Supplementary Information Figure S5). We also performed control experiments by heating $WSe_2$ FETs in vacuum, but we did not observe significant change in device on-state current (Figure S6). This experiment confirms that heating-induced removal of surface impurities on $WSe_2$ does not play an important role in improving $WSe_2$ device performance upon air heating.

The performance of $WSe_2$ FET can be further enhanced by combining the oxidation method developed in this work with other well documented methods, for example, supporting 2D materials on flat and inert BN substrates. As an excellent dielectric material, BN has been reported to have the capability of providing dielectric interfaces with very little charge impurities to graphene, TMDCs, and other 2D materials. The schematic of a back-gated FET using $WSe_2$ on BN is shown in Figure 5a. First we exfoliated BN flakes on $Si/SiO_2$ substrates using a scotch tape. Then, monolayer $WSe_2$ flakes were grown directly on BN using CVD method described above. PL spectrum confirmed that this $WSe_2$ flake on BN is a monolayer (Figure 5b). The successful growth of monolayer $WSe_2$ on BN flakes allows us to fabricate FETs based on this kind of vertical heterostructures. The height of exfoliated BN layer was measured by AFM to be 317 nm (Supplementary Information Figure S7). From the transfer characteristics, again we observed improvement of on-state current and field effect mobility after 250 °C air heating of $WSe_2$ FET on BN substrates (Figures 5c-5d). Notably this monolayer $WSe_2$ device showed a room temperature mobility of 83 $cm^2 \cdot V^{-1} s^{-1}$, which is much higher than devices on $Si/SiO_2$ substrates (Figure 1). The output characteristics ($V_{ds}$-$I_{ds}$) and $V_{ds}$-$I_g$ family curves are shown in Figure 5e-5f, showing nice transistor behavior of FETs made from $WSe_2$ supported on BN substrates. These results suggest that the oxidation method can be combined with other fabrication techniques in enhancing the mobility of $WSe_2$ FETs even further.

**CONCLUSIONS**

In conclusion, we have demonstrated that air heating of $WSe_2$ FETs resulted in a significantly improved device performance *via* the formation of in-plane heterojunctions of $WSe_2$ and their oxidation species $WO_{3-x}$, which is highly conductive. After optimized heating at 250 °C, CVD monolayer $WSe_2$ transistors showed a maximum hole mobility of 51 cm$^2$/V s, an average mobility of 31 cm$^2$/V s, and the device on/off ratios reached 5x10$^8$. We further showed that the same method also works for PVD-grown few-layer WSe$_2$ transistors, demonstrating the generality of the air heating method in improving the FET performance of WSe$_2$. Spectroscopic and TEM characterization showed the formation of WO$_{3-x}$ after air heating, which forms in-plane quilted heterostructures with WSe$_2$ matrix and in turn resulted in an improved transistor performance of p-type WSe$_2$. We also demonstrated that combinations of oxidation method developed in this work with supporting WSe$_2$ flakes on BN substrates further pushed the device performance to a higher level. Our results suggest that controlled oxidation may become a general approach for enhancing performance of FETs made of WSe$_2$ and potentially many other TDMCs.

## METHODS

**CVD growth of monolayer WSe$_2$ flakes.** WO$_3$ (260 mg, 99.9%, Sigma Aldrich) and Se (440 mg, 99.5%, Sigma Aldrich) powders were used as tungsten and selenium source for CVD growth of monolayer WSe$_2$, as reported in our previous work.[19] A three zone furnace was used where Se powders were put in the first zone, and WO$_3$ powders were put in the third zone. The distance between the two sources was 55 cm. The temperatures of WO$_3$ and Se were 950 and 540 °C, respectively. The growth substrates were Si/SiO$_2$ (300 nm) and WSe$_2$ growth was conducted at 950 °C under ambient pressure and Ar/H$_2$ flow rates of 320/20 sccm for 15 min. After growth, the furnace was cooled down to room temperature naturally under protection of Ar/H$_2$ (320/20 sccm).

For the growth of monolayer WSe$_2$ on BN, we first exfoliated BN flakes on Si/SiO$_2$ substrates using a scotch tape. After cleaning the BN flakes, the substrates were loaded into the three zone furnace mentioned above. Then the growth of WSe$_2$ was conducted at 950 °C under ambient pressure and Ar/H$_2$ flow rates of 200/20 sccm for 20 min. We found that the growth of WSe$_2$ was slower and growth yield was lower on BN substrates than direct growth on Si/SiO$_2$ substrates.

**PVD growth of few-layer WSe$_2$ flakes.** WSe$_2$ powders were used as source material for the PVD synthesis of few-layer WSe$_2$ flakes. The growth temperature was 900 °C and the pressure was 700 mTorr during the PVD growth process. A small amount of Ar (10 sccm) was introduced during growth, and the growth lasted for 15 minutes.

**Materials characterization.** The thickness and layer numbers of CVD and PVD-grown WSe$_2$ flakes were evaluated using AFM (Dimensional 3100, Digital Instrument, tapping mode), Raman spectroscopy and PL spectroscopy (Renishaw Raman with a 532 nm excitation laser and a laser spot size of ~1 μm). The air heated WSe$_2$ flakes were characterized using Raman, XPS (The Kratos Axis Ultra DLD surface analysis instrument using focused monochromatized Al Kα radiation), XRD (Rigaku Ultima IV powder/thin-film diffractometer with Cu Kα line), and TEM (JEOL 2100F, 200 kV).

**Device fabrication and measurements.** E-beam lithography was used to fabricate back gate WSe$_2$ devices. A bilayer PMMA e-beam resist was spin-coated on substrates and EBL was conducted to defined the source and drain electrodes. Ti/Pd (0.5 nm/ 50 nm) was deposited as source and drain contacts, and bottom silicon substrate serves as a global back gate. The device measurements were conducted in ambient condition using an Agilent 4156B Semiconductor Parameter Analyzer. Air annealing of $WSe_2$ devices were performed by heating $WSe_2$ devices on a hot plate under lab environment at different temperatures (50 to 400 °C) for 1 hour. As control experiments, vacuum annealing was performed in a tube furnace at 200 °C for 1 hour at a pressure of 12 mTorr. The field-effect mobility of $WSe_2$ FETs was calculated using the following equation:

$$\mu = \frac{L}{W} \frac{1}{C_{ox} V_{ds}} \frac{dI_{ds}}{dV_g}$$

where L and W are the channel length and width of the FET. $V_{ds}$ is the source-drain voltage. $I_{ds}$ is the current flowing from source to drain and $V_g$ is the gate voltage. $C_{ox}$ is the gate capacitance per unit area.

**ASSOCIATED CONTENT.**

**Supporting Information.** The Supporting Information is available free of charge on the ACS Publications website at DOI: nn-2016-00527g.

Additional AFM, Raman, PL, Optical microscopy, and device measurement results (PDF).

**Acknowledgements**. The authors acknowledge Keji Lai and Di Wu of UT Austin for helpful discussions. This work was supported by the Air Force Office of Scientific Research (AFOSR). We would like to acknowledge the collaboration of this research with King Abdulaziz City for Science and Technology (KACST) *via* the Center of Excellence for Nanotechnologies (CEGN).

**Figure Legends.**

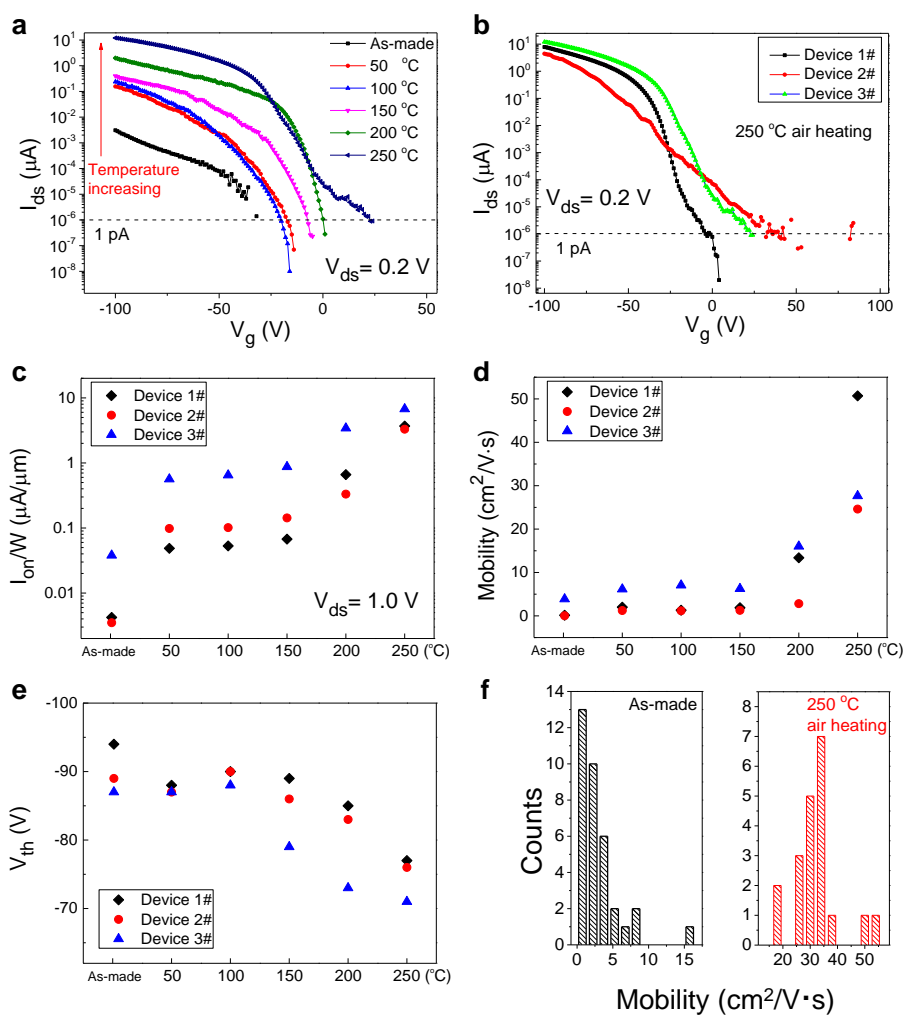

**Figure 1.** Improved FET performance of CVD-grown monolayer WSe$_2$ *via* controlled heating in air. (a) I$_{ds}$-V$_g$ curves of a representative monolayer WSe$_2$ FET in as-made state as well as after air heating at different temperatures. (b) I$_{ds}$-V$_g$ curves of three WSe$_2$ transistors after 250 °C air heating. (c), (d), (e) Changes of I$_{on}$/W (c), field-effect mobility (d), and V$_{th}$ (e) *versus* heating temperatures for three devices shown in (b). (f) Mobility distributions of monolayer WSe$_2$ FETs in as-made state and the same devices after 250 °C air heating.

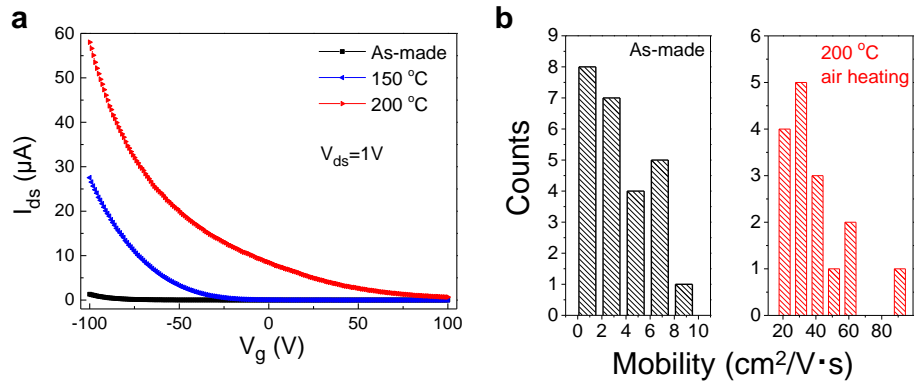

**Figure 2.** Improved FET performance of PVD-grown few-layer WSe$_2$ FETs *via* controlled heating in air. (a) I$_{ds}$-V$_g$ curves of a representative few-layer WSe$_2$ devices at different conditions. (b) Field effect mobility distributions of few-layer WSe$_2$ FETs in as-made state and after 200 °C air heating.

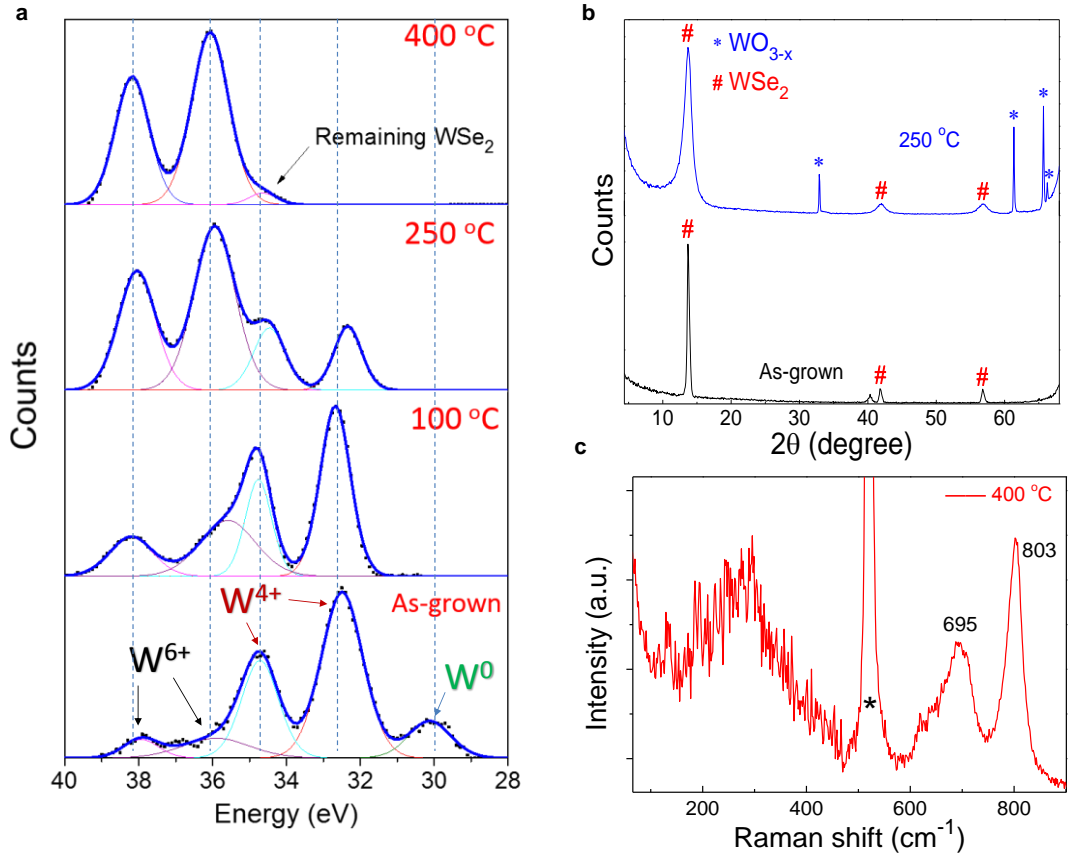

**Figure 3.** Spectroscopy studies of $WSe_2$ samples. (a) XPS spectra of as-grown few-layer $WSe_2$ sample and the same sample after air heating at different temperatures. Note that the intensity ratios of $W^{6+}/W^{4+}$ increase with increasing heating temperatures. (b) XRD patterns of as-grown few-layer $WSe_2$ and the same sample after 250 °C air heating. (c) Raman spectra of PVD few-layer $WSe_2$ sample after air heating at 400 °C. The two peaks at 695 and 803 $cm^{-1}$ can be assigned to tungsten oxide. The peak with a star (*) at 520 $cm^{-1}$ indicates the characteristic peak of Si substrate.

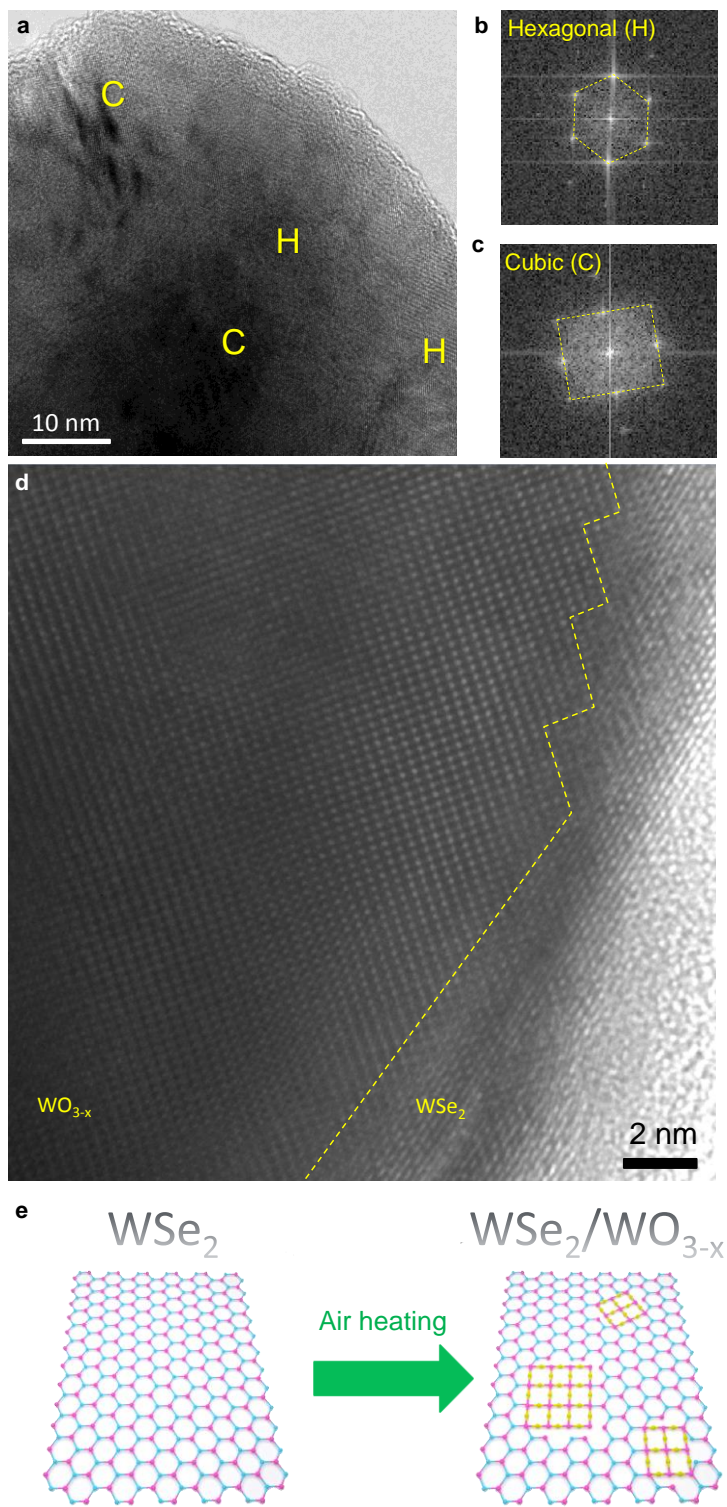

**Figure 4.** TEM characterization and schematics of in-plane heterojunctions of WO$_{3-x}$ (cubic) and WSe$_2$ (hexagonal). (a) A typical TEM image of WSe$_2$ sample after 150 °C air heating. Letter C stands for cubic while letter H stands for hexgonal. (b) FFT taken

at the letter H position, showing hexagonal WSe$_2$ lattice. (c) FFT taken at the letter C position, showing cubic WO$_{3-x}$ lattice. (d) A HRTEM image showing the interface (yellow dash lines) between WO$_{3-x}$ and WSe$_2$ domains. (e) Schematics of structural change during oxidation of WSe$_2$ during air heating, showing the formation of in-plane quilted heterostructures made of WSe$_2$ and WO$_{3-x}$.

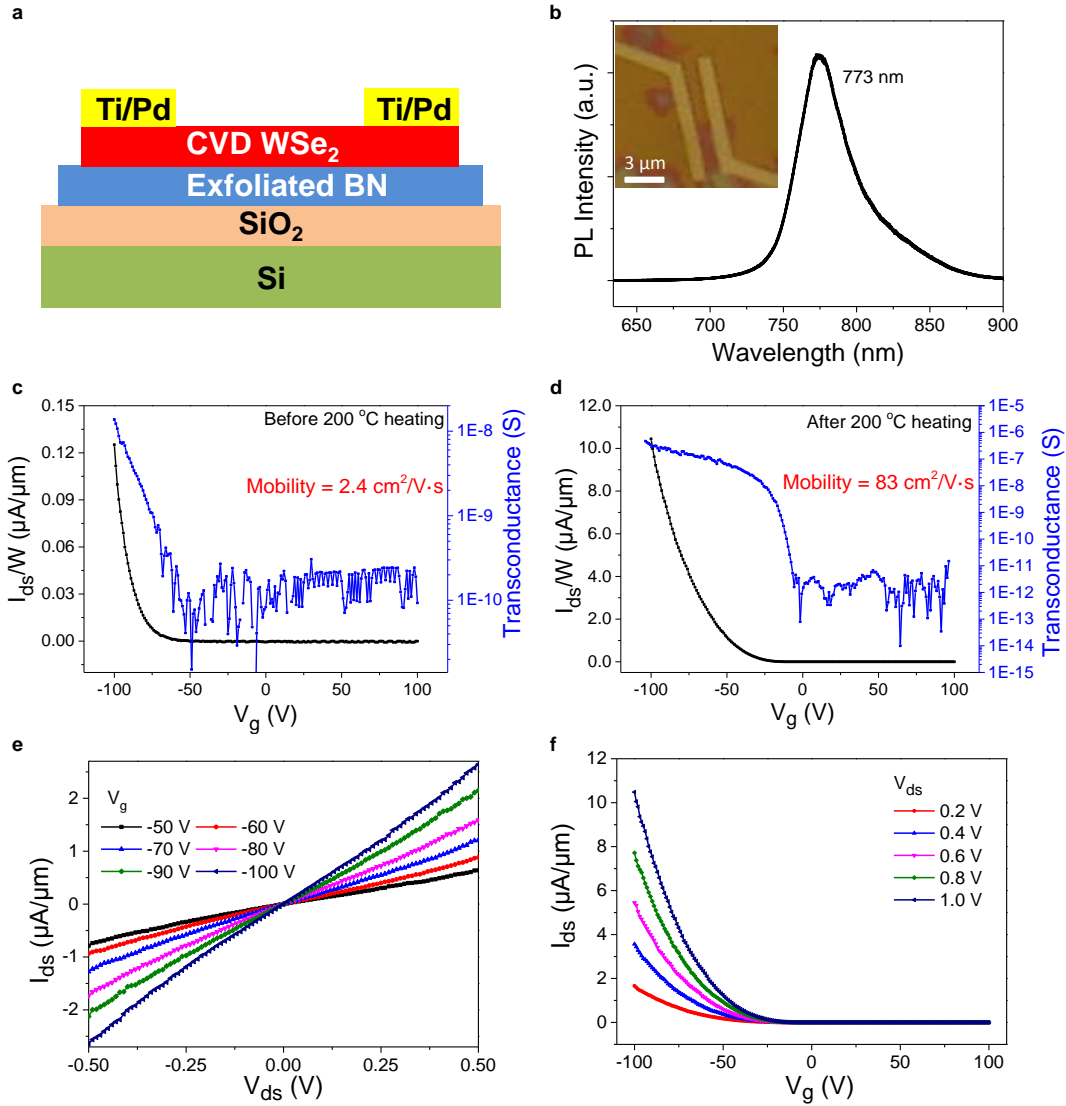

**Figure 5.** WSe$_2$ on BN devices. (a) Schematic of a FET based on WSe$_2$ grown on BN. (b) PL spectrum of the monolayer WSe$_2$ grown on BN. Inset shows optical image of a FET device with WSe$_2$ on BN flake. (c), (d) I$_{ds}$-V$_g$ curves of the device shown in (b) before and after air heating at 200 °C. (e), (f) I$_{ds}$-V$_{ds}$ and I$_{ds}$-V$_g$ family curves of the device shown in (b) after air heating at 200 °C.

TOC Figure

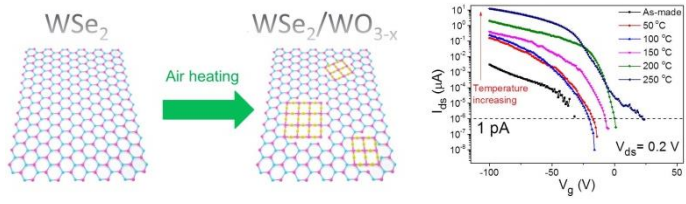